# Nanomechanics of MXene flakes at gold interfaces

*Pierluigi Bilotto[1]*, Sabine Schwarz[2], Marko Piljevic[3], Iago Peters[4], Michael Stöger-Pollach[2], Carsten Gachot[1]*

[1] Institute of Engineering Design and Product Development, Research Unit of Tribology (E307-05), TU Wien, Lehargasse 6, Vienna, 1060, Austria

[2] University Service Center for Transmission Electron Microscopy, TU Wien, Stadionallee 2, 1020, Vienna, Austria

[3] AC2T research GmbH, Viktor-Kaplan-Straße 2/C, Wiener Neustadt, 2700, Austria

[4] Institute of Applied Physics, E134-02, TU Wien, 1040, Vienna, Austria, Wiedner Hauptstrasse 8, Vienna, 1040, Austria

ABSTRACT

The demand for novel self-powering and sustainable electronics requires major efforts in identifying new advanced materials for nano-applications. MXene have gathered attention due to their electronic and mechanical properties, however, their nanomechanical compliance against a metal-like interface is still not clearly identified. In this work, we employ atomic force microscopy to characterize the nanomechanical properties of a self-assembled thin flake of titanium carbide

MXene ($Ti_3C_2T_x$) against a gold probe. The investigation returns an interfacial shear stress of 399 MPa, and the observation of nano-wear localized in the center of the MXene flake. Nevertheless, MXene flakes retained their crystallinity in the tribofilms as confirmed by transmission electron microscopy and electron energy loss spectroscopy. The outcomes of this work set the basis for the use of $Ti_3C_2T_x$ in novel nano harvesting systems involving metal interfaces (e.g., tribovoltaic nanogenerators) with large scope in nanoelectronics, wearable sensing, electric vehicles, and robotics.

## 1 Introduction

The discovery of graphene in 2004 revolutionized materials science by revealing that dimensionality profoundly influences chemical and physical properties.[1] MXenes, 2D transition metal carbides, nitrides, and carbonitrides first reported in 2011,[2] have since gained attention for their exceptional electrical, optical, and mechanical properties.[3] MXenes emerge in comparison to other 2D materials because they are a family of materials on their own, allowing complex to highly entropic compositions.[4] Moreover, surface terminations and defects at MXene interfaces have been addressed as key actors in dictating physical and chemical properties, ranging from catalysis to mechanical and tribological performances.[5–7]

Titanium carbide ($Ti_3C_2T_x$) is the most studied and currently understood MXene. It has been tested in terms of mechanical and tribological due to its ability to act as a solid lubricant with large scope in reducing friction and wear in extreme conditions.[8,9]

The lubrication properties of MXenes derive from the complex interplay of phenomena taking place at the flakes' interface, resulting in the formation of a stable tribofilm. This includes the probability of adhering to the counterbodies,[10] the activation of a reservoir mechanism to maintain flakes in the contact region,[11] the presence of defects affecting friction at the nanoscale, the geometrical distribution which defines the coating quality, sacrificial layering (e.g., in hybrid MXenes) within the tribofilms composition,[12] and environmental characteristics (e.g., humidity, load, temperature, speed). All these properties originate from the nanoscale, driving researchers to extensively study the Nanomechanics of MXenes with simulation and experimental tools such as Density Functional Theory (DFT), Molecular Dynamics (MD), and Atomic Force Microscopy (AFM). [7,10,13–15]

The current state of the art presents well-documented information on how single MXene flakes should behave at the nanoscale, which includes adhesion and nanoindentation tests providing work of adhesion and Young modulus of $Ti_3C_2T_x$,[16] tensile fracture of monolayer $Ti_3C_2T_x$,[17] and nanofriction investigation.[18–20] Nevertheless, the research activity has not extensively studied a critical parameter that encompasses the complexity of friction at the nanoscale, that is the interfacial shear stress.

Only recently, different interfaces involving two-dimensional materials (2DM) such as $Ti_3C_2T_x/Ti_3C_2T_x$, $Ti_3C_2T_x$/graphene, and $Ti_3C_2T_x/MoS_2$ were investigated, delivering an interfacial shear stress below 2 MPa.[15] Still, the experiment was performed with a silica sphere on a tipless cantilever, a counterpart similar to previous nanomechanical investigations involving Si or $Si_3N_4$ tips.[13,18] Such an experimental design is important for standardization and to effectively mimic a dielectric/two-dimensional material interface. However, the interest for nanotribological compliant 2DM interfaces is rising in nanoelectronics due to the direct effect between friction and the electron charge density (e.g., graphene FET devices).[21]

The term triboelectric nanogenerators (TENG) encompasses all those materials that are able to combine the effects of contact electrification and electrostatic induction to convert harvested mechanical energy into an electric signal.[22] TENGs have revolutionized nanoelectronics, finding application up to the macroscale and daily life (e.g., machine monitoring, wearable sensors).[23–25] More recently, a new physical phenomenon has been discussed where a direct current is generated by allowing the creation of an electron-hole pair. The tribovoltaic nanogenerator (TVNG) effect offers a novel perspective; however, it imposes the use of metal/metal-like or semiconductor-like sliding interfaces.[26–28] This limitation offers an important window of interest for MXenes, as their chemical tunability (from terminations, element compositions, or defect control) allows them to tune their physical properties to express semiconductor- and/or metal-like characteristics.[29,30]

For instance, the TVNG effect in a MXene–silicon heterojunction has recently been investigated for self-powered sensing applications.[31] In that work, MXene powder was compacted into a solid pellet (MXene slider) designed to slide against a silicon wafer. This approach enabled measurement of key performance parameters such as a peak output current of up to 22 μA under a normal force of 4.56 N. However, compacting the MXene into a rigid slider constrained the degrees of freedom of individual MXene flakes during sliding, thereby compromising the nanoscale tribological behavior (i.e., solid lubrication) that is essential for next-generation miniaturized nanoelectronics. It is therefore essential to characterize the nanomechanical performance of individual MXene flakes against metallic probes, providing a mechanistic basis for their tribological behavior.

Hereby, we explore for the first time the nanomechanical performance of isolated MXene flakes ($Ti_3C_2T_x$) probed by a metal counterpart (i.e., gold - Au) by Atomic Force Microscopy (AFM). $Ti_3C_2T_x$ flakes deposited by interfacial self-assembly (ISA-MXene) were characterized by Force Spectroscopy (FS-AFM) and Lateral Scanning (LS-AFM) AFM; then, they were investigated by Transmission Electron Microscopy (TEM) to confirm the resilience and crystallinity of the flakes after nanomechanical stress. Finally, we detailed the spatial evolution of friction force $F_f(x, y)$, adhesion force $F_a(x, y)$, and interfacial shear stress $\tau(x, y)$ by force mapping AFM (FM-AFM) analysis.

We obtained an average $\tau_{MXene} = 399 \pm 15$ MPa for a single ISA-MXene flake on mica against gold. We critically addressed the results, to be strongly dependent on the specific metal-MXene interfaces as Au largely affects the overall intermolecular forces. Consequently, this approach is instrumental in mimicking scenarios of MXene interacting with metal surfaces in novel energy harvesting and self-powering electronics, biosensing, wearable electronics, or similarly in predicting how MXene interfaces would interact with metal-like wear particles in nanoconfined systems.

## 2. Results and discussion

Isolated MXene flakes were prepared on freshly cleaved mica substrates by interfacial assembly with six times immersions (more details in the experimental section), as illustrated in **Figure 1a**.[32] The as-prepared interfacial self-assembled MXene (ISA-MXene) was then transferred into the AFM chamber were topography confirmed the presence of large-scale ultrathin films (see Figure S1).

### 2.1 Topography and film thickness of ISA-MXene

**Figure 1b** shows the topography of freshly deposited ISA-MXene, i.e., $Ti_3C_2T_x$, transferred onto the AFM and imaged at room temperature conditions ($\sim 25$ °C). Eleven clearly defined flakes (F) are identified in a 14μm × 14 μm area. Statistics on F (**Figure 1c**) show an average height $\langle H \rangle = 1.25 \pm 0.61$ nm and an average width of $\langle w \rangle = 1.37 \pm 0.64$ $\mu$m. The flake F.7 appears to be larger than the others with a width of $w_{F.7} \sim 3$ $\mu$m (see **Figure 1d**). Nevertheless, the measured thickness is in line with the other flakes (e.g., $T_{F.7} = 1.24 \pm 0.14$ nm along the green section in **Figure 1d**), ruling out the possibility of having strong clustering effects and defect nucleation in the direction orthogonal to the plane. Additional large scans on different areas of the sample confirmed the homogeneous thickness of ISA-MXene on mica (see **Figure S1**). Surface Forces Apparatus (SFA) is used for the first time to evaluate the thickness of an MXene film and corroborate the AFM topography results. At the contact position (more info in the experimental section), we measured a film thickness of $T_{SFA} = 1.5 \pm 0.1$ nm for ISA-MXene (see **Figure S2**), which is indeed in line with the AFM results.

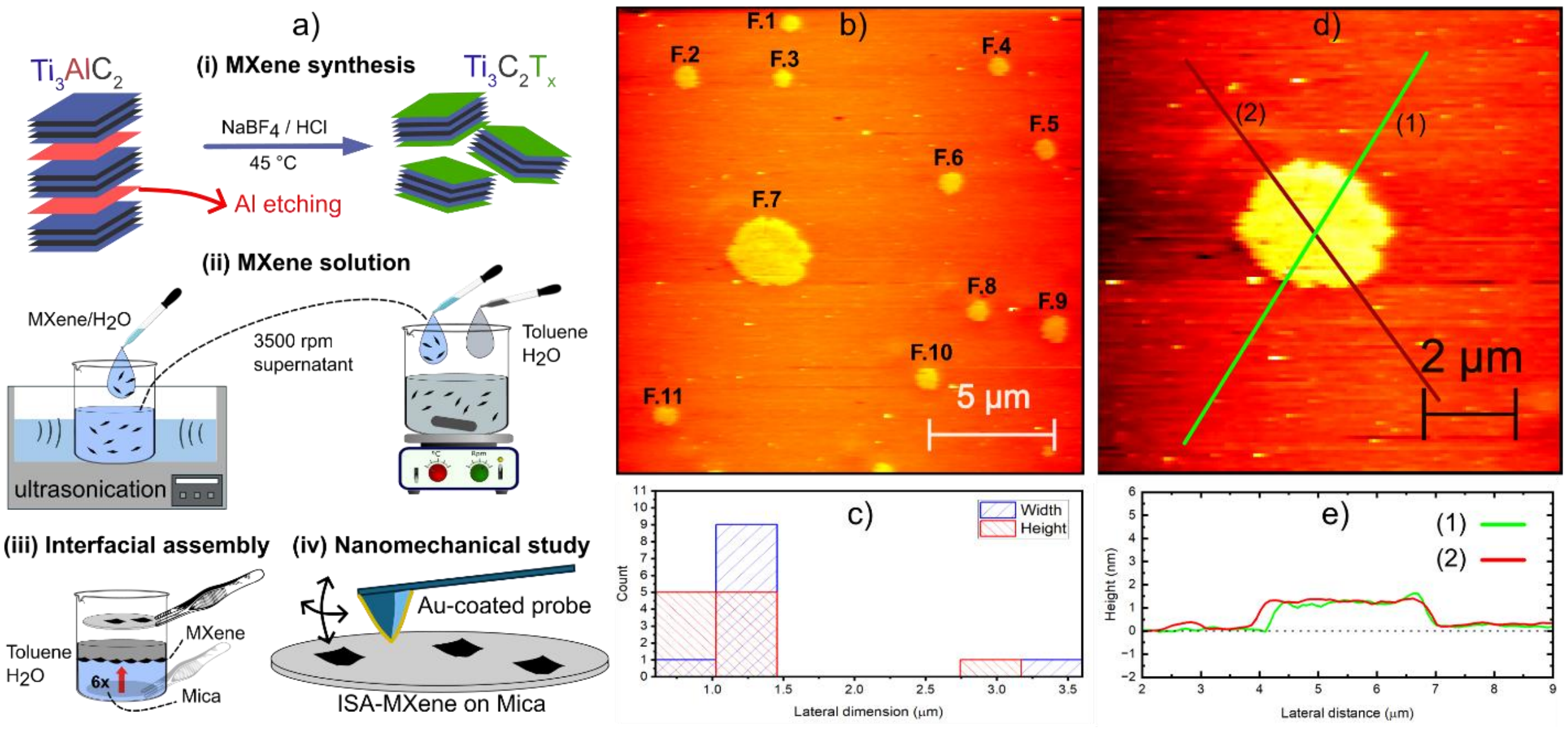

*Figure 1: a) Preparation of interfacial self-assembled MXene (ISA-MXene) flakes on mica: (i) synthesis, (ii) dispersion, (iii) assembly, and (iv) AFM experimental setup. b-e) AFM topography of the ISA-MXene flakes on mica. b) Statistics on the flakes: 11 flakes (F) are counted, with values of height and width shown in panel c). The average width is about 1.2 µm. d) Zoom into F.7. e) Profile of the sections shown in panel d. The measured average thickness of the ISA-MXene flake is 1.24 ± 0.14 nm (green line) and 1.29 ± 0.06 nm (red line).*

It is worth noting that the flakes' topography appears extremely stable over time. **Figure S3** shows the topography of the same sample acquired after four months. The sample was stored in a Petri dish in the dark at room temperature (no vacuum or inert atmosphere). Figure S3 shows that these isolated flakes kept their circular shape and average thickness within the error bar. Thus, ISA-MXene flakes appear to be extra thin yet resilient, making them ideal for nanomechanical investigation.

**2.2 Adhesion study on an isolated ISA-MXene flake.**

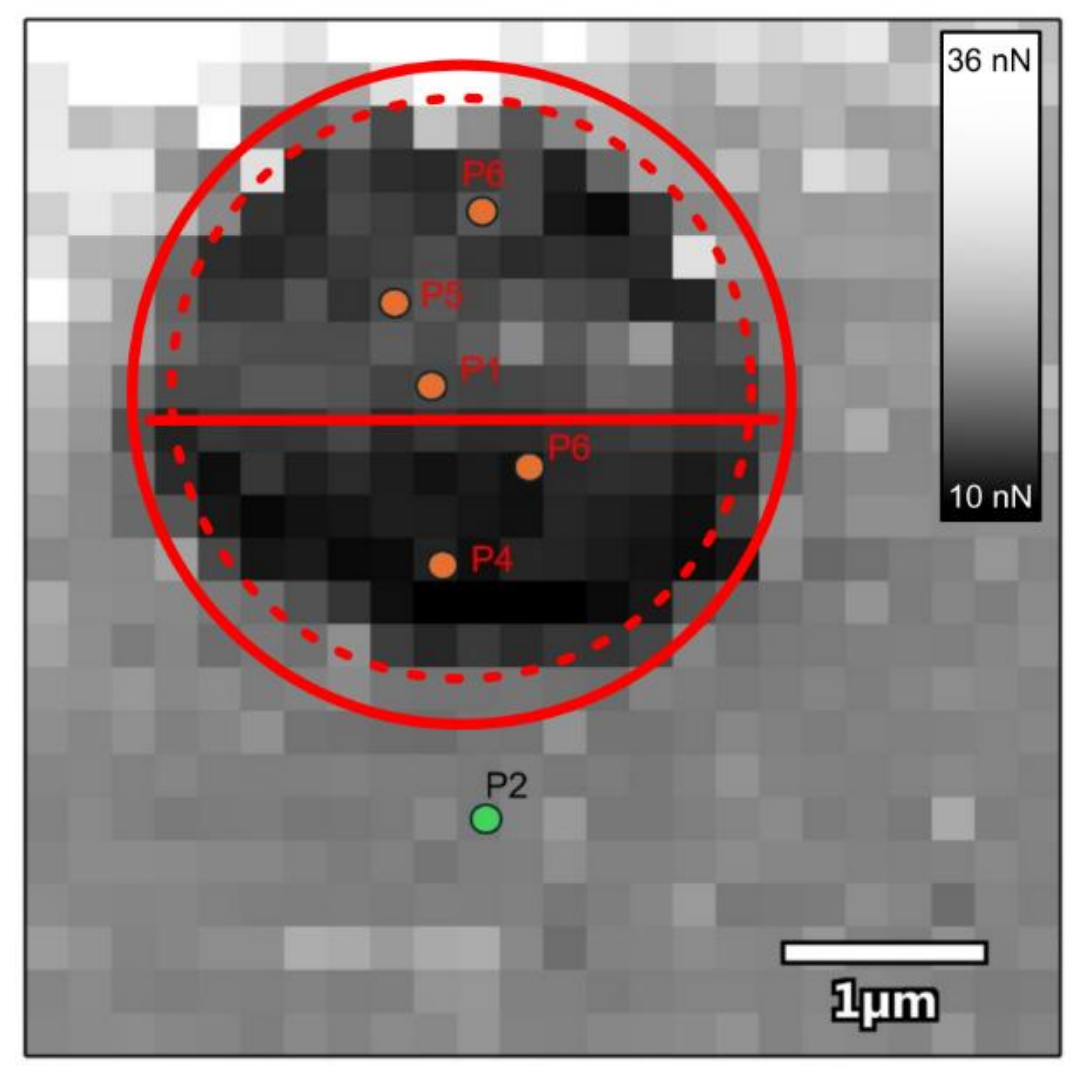

*Figure 2: Adhesion Map (FM1) of a MXene flake on mica substrate measured in air. Six points are selected to perform further nanomechanical characterization.*

Force Spectroscopy Atomic Force Microscopy (FS-AFM) was conducted to extract the average adhesion force of F.7 (Figure 1).

**Figure 2** shows the adhesion map (FM1) on F.7. The reconstructed FS-AFM image presents an adhesion distribution that follows the topographical detail of F.7 (see Figure 1d).

Then, the topographical image (Height signal) obtained while acquiring FM1 (see **Figure S4**) was used to characterize the dimensions of the flake's perimeter appearing in the adhesion map. In detail, the two red circumferences in Figure 2 indicate the minimum (dashed line) and maximum (full line) surface occupied by the reconstructed image. Thus, the image obtained from FS-AFM has lateral dimensions within the area of a circle with diameter $\delta = 3.0 \pm 0.2$ µm (indicated by the red flat line), in line with topographical results (Figure 1 and Figure S4). The error of $\delta$ corresponds to a single pixel size.

The distribution of adhesion forces measured during the force maps is shown in **Figure S5**. In detail, FM1 shows a two peaks distribution corresponding to mica (Area 1) and ISA-MXene (Area 2), with average adhesion forces of $\langle F_a \rangle_{A1} = 29.6 \pm 3.8\,\text{nN}$ and $\langle F_a \rangle_{A2} = 11.2 \pm 1.1\,\text{nN}$ respectively for mica and $Ti_3C_2T_x$.

The former value is expected given a gold-coated tip interacting against mica in ambient conditions, where capillary forces (due to mica's hydrophilicity) contribute to short-range interactions. The decrease in average adhesion found on the flake (i.e., $\langle F_a \rangle_{A2}$) is related to the interfacial properties of $Ti_3C_2T_x$.

First, the ISA-MXene flakes present a distribution of F- terminations which lower the hydrophilicity, namely the capillary force contribution to the pull-off force. Moreover, by defining the work of adhesion as $W_a = \langle F_a \rangle / \lambda R_{tip}$ where $R_{tip}$ is the nominal tip radius and $\lambda = 1.560$ a fitting parameter for MXenes, we obtain $W_{MXene} \sim 0.7\ \mathrm{mJ/m^2}$, which is in accordance with nanomechanical experiments and simulations of a single layer $Ti_3C_2T_x$ flake.[18,19,33,34]

For comparison, a force map was performed on a different flake (Area 3). The average adhesion $\langle F_a \rangle_{A3} = 11.6 \pm 1.9$ nN is in accordance with $\langle F_a \rangle_{A2}$, confirming homogeneous nanomechanical behavior between the flakes (see Figure S5).

**2.3 Derivation of interfacial shear stress $\tau$**

Lateral Force AFM (LF-AFM) was used to evaluate the nanotribological performance of ISA-MXene (see **Figure S6** for clear contrast between flake and substrate during LF-AFM). Unprocessed LF-AFM scans showed a shadowing effect; however, by measuring the positions of shadows and reference objects in consecutive runs, the possibility of a continuous drag phenomenon (see **Figure S7**) was ruled out. In other words, the flakes might have been dislodged in the very first tip scan but then remained fixed in subsequent scans, suggesting that there was no relative motion between the flake and the substrate during scanning.

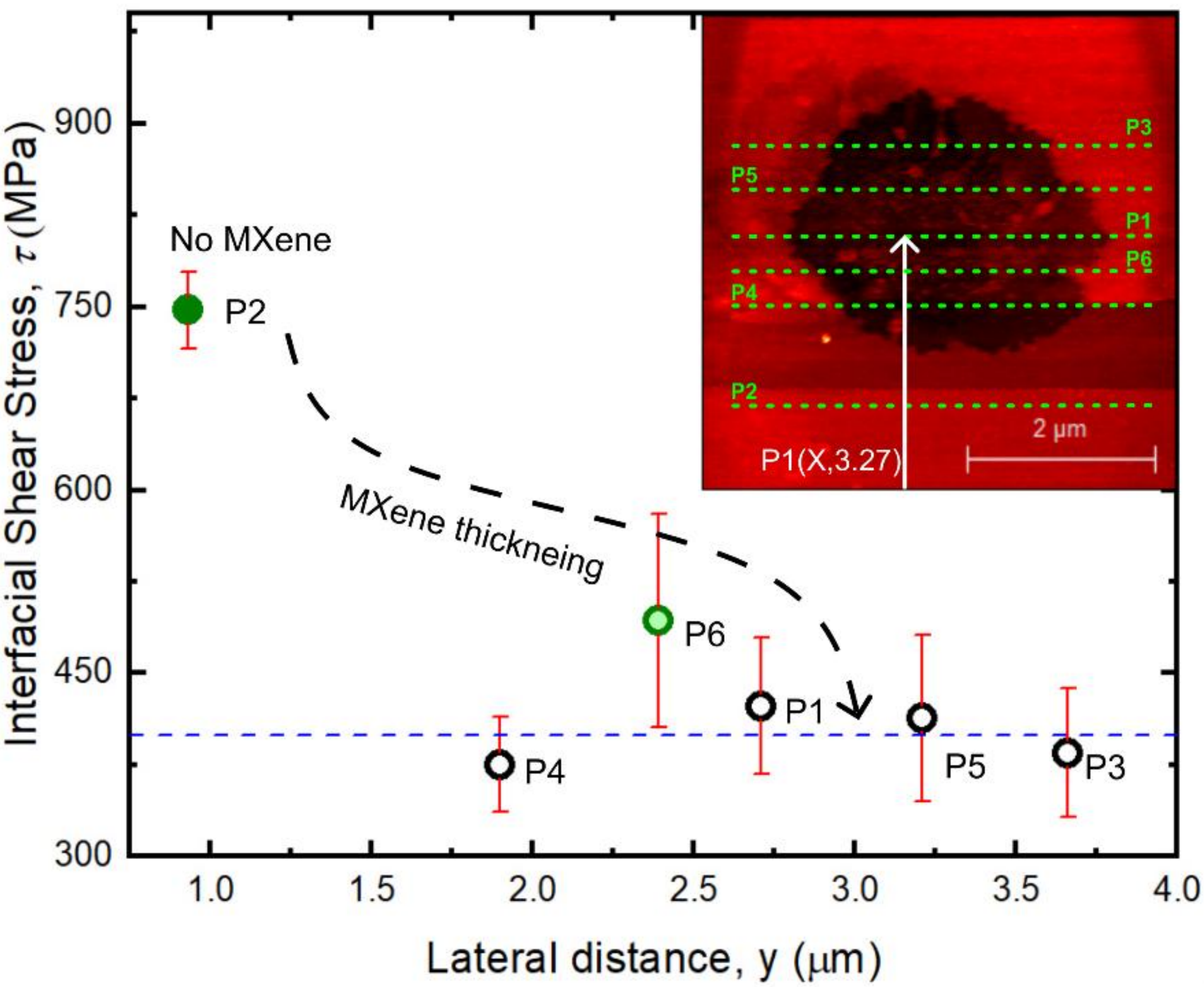

*Figure 3: Evolution of the interfacial shear stress (τ) across an ISA-MXene flake parametrized by six points (P) and their lateral distance coordinate y. The location of the points and the flake are visualized in the friction map shown as inset. The full green point corresponds to an interaction with the mica substrate while, the open circles to a τ mediated by the MXene presence. The open light green point corresponds to a part of the flake highly damaged by nano wear. The blue dashed line indicates the weighted average value of τ across the ISA-MXene area.*

Given a fixed load at $F_N = 64.95$ nN, we obtained a friction map of F.7 by arithmetic operation on lateral trace and retrace signals as shown in the inset of **Figure 3** (more details in the Experimental section). Six points Pn(x,y) (with n= 1,…,6) were defined on the adhesion map illustrated in Figure 2. From those, we drew sections perpendicular to their y coordinate, which correspond to the sliding distance of friction curves (each line of the friction map is a friction curve). The explicit friction curves per Pn(x,y) are presented in **Figure S8**.

The weighted average of the friction forces for each point Pn(x,y) onto the MXene flake is $\langle F_f \rangle_{MXene} = 8.19 \pm 0.57$ nN, where the error is the standard deviation.

Friction forces at the nanoscale are modulated by the real contact area $A(x,y)$ and adhesion forces rather than normal load $F_N$ only (Amontons´ law).[35] Given the elastic constant of the AFM cantilever and the measured low adhesion values (~10 nN), we define the real contact area *A(x,y)* according to the Derjaguin-Muller-Toporov model (DMT):[36]

$$A(x,y) = \pi \left[ \frac{3R^*}{4\, E^*_{\mathrm{Ti3C2Tx}}} * \left( F_N + F_a(x,y) \right) \right]^{2/3}$$

where $E^*_{\mathrm{Ti3C2Tx}}$ is the effective elastic modulus at the MXene/Au interface (more details in the experimental section), and $R^*$ is the approximated effective radius of curvature of the AFM tip (see Expetimental section). Then, given the real contact area and friction forces, we define the interfacial shear stress in (x,y) as:

$$\tau(x,y) = \frac{F_f(x,y)}{\pi\left[\frac{3R^*}{4\,E^*_{\mathrm{Ti3C2Tx}}} * \left(F_N + F_a(x,y)\right)\right]^{2/3}}$$

**Figure 3** shows the spatial parametrization of $\tau(x,y)$ along the lateral distance y, revealing the role of intermolecular forces in dictating interfacial shear stress of the scanned area (mica + ISA-MXene).

First, $\tau(x,y)_{P2}$ shows the highest value, which corresponds to the direct interaction between the AFM gold tip and the hydrophilic mica substrate.

Second, an adhesion damping effect in the interfacial shear stress is observed within the flake perimeter due to the strong correlation between adhesion and $\tau(x,y)$.

Third, the interfacial shear is homogeneous across the flake within the error values, allowing for mechanical differentiation between ISA-MXene and the substrate.

The measured value for the Au/mica interface at ~750 MPa is expected given the ambient conditions, capillary contributions, and DMT modelling of the contact mechanics.[37]

As gold adheres strongly to mica, when the tip interacts with the ISA-MXene flake, the Van der Waals contribution between mica and gold is modulated by the flakes' thickness, reducing the adhesion forces as confirmed by force mapping (see Figure 2 and arrow in black dashed arrow in Figure 3).

It is worth noting that surface terminations of MXenes, which typically modulate adhesion and friction as discussed in experiments and DFT simulations,[10,34] might play a less relevant role at the investigated interface given the low chemical reactivity of gold. Finally, the ambient conditions foster the formation of capillary bridges, further damping the attraction contribution to the mica, resulting in the lower adhesion. Thus, from this adhesion-driven mechanism, real contact area and friction forces are tuned accordingly, resulting in a ~60% reduction in friction forces and a consequent average interfacial shear stress on the ISA-MXene flake of the $\tau_{MXene} = 399$ MPa.

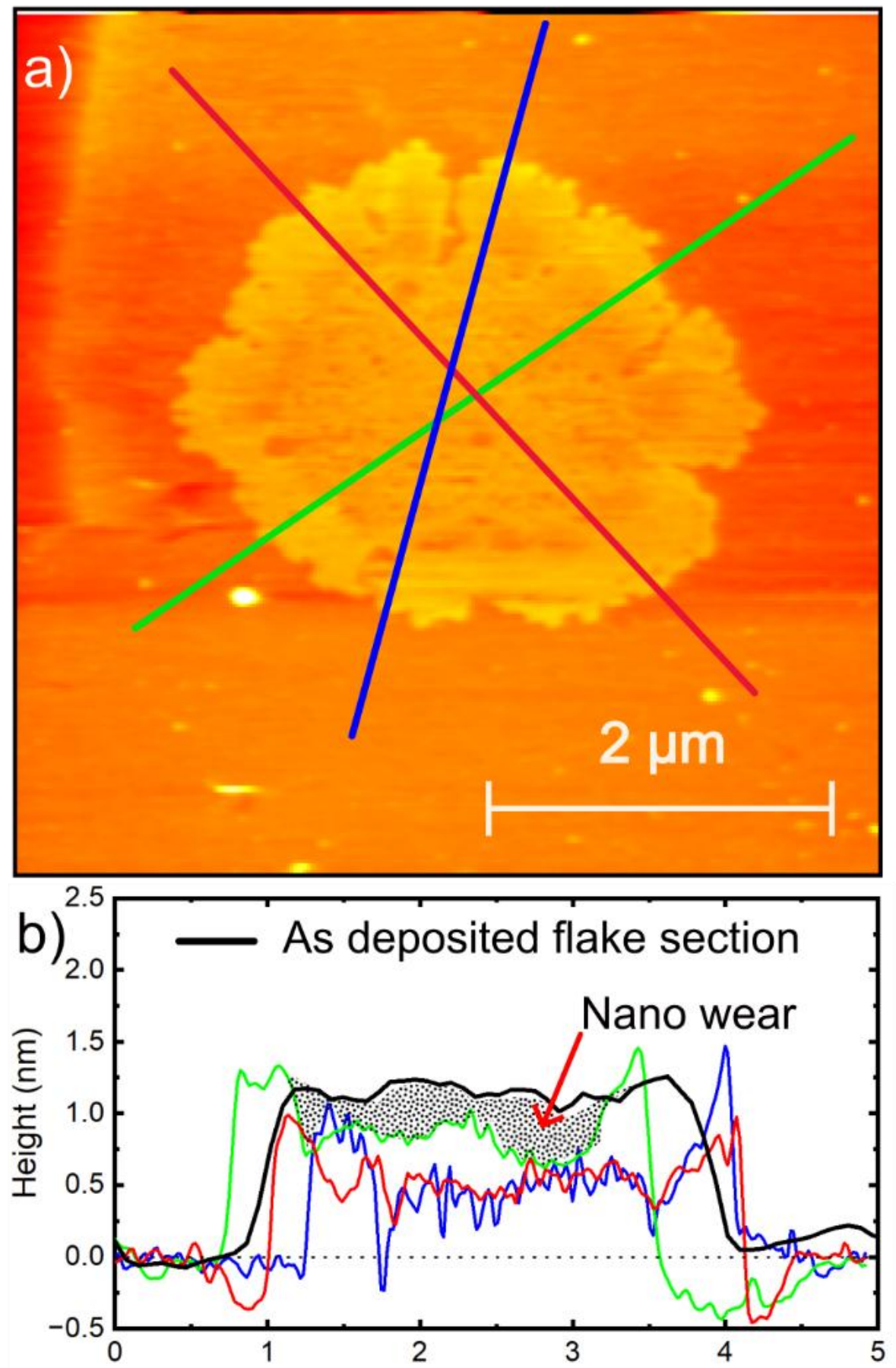

*Figure 4: Topography of F.7 after the nanomechanical test. a) High resolution topography image and defined sections (blue, red, and green line). b) Line plot of the selected section. In black is one of the sections in Figure 1d used here as a reference. The dotted area shows a visualization of the nano-wear.*

To the best of our knowledge, this is the first reported interfacial shear stress at the Au/$Ti_3C_2T_x$ interface. The current state of the art is scarce in results on the interface between gold and 2D materials for direct comparisons. For instance, the interaction between $MoS_2$ monolayers against and Au-coated tip showed an $\tau$ = 130 MPa,[38] while the interfacial shear stress involving a gold/hexagonal boron nitride interface has been reported as $\tau$ ~ 700 MPa; however, the Au probe was a slab manipulated by the AFM tip, not the actual coated tip.[39]

High-resolution AFM topography was conducted on F.7 following the nanomechanical tests to check for structural resilience. **Figure 4** shows the AFM image of F.7 after nanomechanical stress, revealing (Figure 4a) that the perimeter and overview shape of the ISA-MXene flake is maintained (see inset Figure 3 for comparison and Figure 1d). However, the section profiles (Figure 4b) present a more damaged surface than the one discussed in Figure 1d (reference black height in Figure 4b). We interpret this as damage occurring at the nanoscale (e.g., nano wear) resulting from the metal-MXene interaction in air (see dotted area in Figure 4b). The fact that nanoscopic wear is more accentuated in the centre of the flake might suggest that edges have a better mechanical compliance to the sliding probe. Given the good interfacial shear stress discussed in Figure 3 and the appearance of nano-wear found in Figure 4, it is important to draw a line on whether ISA-MXene are indeed performing well in terms of Nanomechanics at metal contact. Thus, the structural integrity of the flake after nanomechanical tests was evaluated by transmission electron microscopy (TEM).

**2.4 Structural integrity of the flake after the nanotribological test**

TEM was conducted on a Focused Ion Beam (FIB) lamella of ISA-MXene on mica after nanomechanical tests. cross-section scan of the lamella by Energy Dispersive X-ray Spectroscopy (EDX) reveals the presence of elements characteristics of ISA-MXene composition, namely Ti, C, and O (see **Figure S9**). Further details are presented in terms of high-resolution TEM (HR-TEM) in **Figure 5**. Figure 5a shows one of the HRTEM patterns from which we calculate the interlayer spacing distance $d$ (see the inset highlighted in red). In the blue-highlighted insert of Figure 5a, we can further visualize the relative position of the Ti-Ti bonds along the Ti-plane ($x_1$) or across the C layer ($x_2$). The averaged measured distances extracted from different positions (min. 3) in the lamella are: $x_1 = 3.19 \pm 0.04$ Å, $x_2 = 2.72 \pm 0.07$ Å.

These values are in accordance with Ti-Ti distances discussed in literature (from EXAFS measurements or DFT calculations).[40,41] Specifically, the in-plane and out-of-plane (across the C layer) Ti-Ti distance is $x = 3$ Å, fitting well with the reported $x_1$ and $x_2$. The latter appears slightly reduced, possibly due to incomplete relaxation of the bonds following the nanotribological test or the presence of defects.

In Figure 5b, two other different areas are selected from the same lamella to better characterize the interlayer spacing distance. The extracted interlayer spacing distances are in accordance with previous reports on $Ti_3C_2T_x$.[10,12,33,42–44] Small discrepancies in the values for $d$ may be related to different distribution of surface terminations and/or defects at MXene interfaces.

Yellow arrows are used to indicate amorphous regions. In Figures 5a and b, these amorphous layers embed the MXene flakes (or vice versa). The presence of Carbon-like amorphous regions in tribofilms formed by 2D solid lubricants has been observed in many recent works.[12,42,45,46] On one side, the presence of these amorphous regions may have a positive effect on tribological performance by sliding synergistically with two-dimensional flakes. On the other side, they can establish a swelling effect (due to the interdigitation of C) which increases the thickness of the tribofilms, as can be noticed by comparing Figure 5a and 5b.

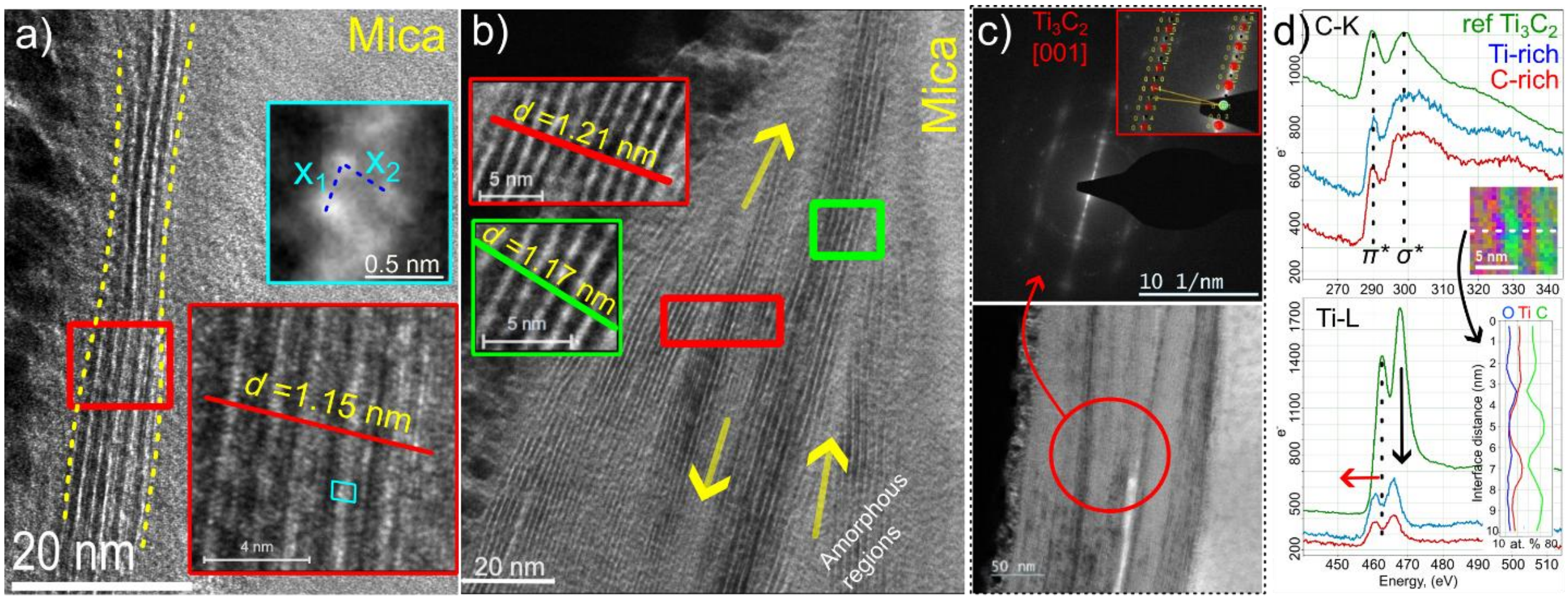

*Figure 5: a) Few-layer flake visualization in a FIB lamella (see yellow dashed lines). The interlayer spacing distance d is calculated by averaging over the bright and dark contrast curves of the layers highlighted in red (see inset). The blue inset shows the measured Ti-Ti distance at the out-of-plane ($x_1$) and across the C layer ($x_2$). b) Different locations of the lamella show that the 2D flakes become embedded by amorphous regions indicated by the yellow arrows. The HRTEM patterns between the amorphous layers reveal the interlayer spacing typical for titanium carbide MXenes. c) Selected area electron diffraction (SAED) pattern showing the $Ti_3C_2$ crystal structure at the [001] plane orientation (calculated pattern shown in the insert). d) ELNES analysis for the C-K and Ti-L edges at the interface with the substrate. In green, a reference spectrum for $Ti_3C_2$ MXene is plotted, while in blue and red curves are plotted Ti and C rich layers of the region of interest. The black downward arrow in Ti-L indicates the reduction of intensity found at the interface, while the leftward red arrow highlights the change in Ti oxidation state. The inset shows an RGB elemental map of the region of interest, with the dashed white line indicating the specific interface on which the elemental composition profile is performed.*

Further confirmation of the titanium carbide nature of the observed flakes comes from selected areas electron diffraction (SAED) and electron energy loss spectroscopy (EELS) results. Figure 5c shows an example of SAED pattern performed in a different area from the ones discussed in Figure 5a and 5b. The diffraction image is fitted with a calculated $Ti_3C_2$ MXene crystal structure at orientation [001], confirming the hypothesis that the tribofilm is indeed constituted of MXene (see Figure S10 for a different pattern of the reference SAED on mica).

Figure 5d shows the Energy Loss Near Edge Structure (ELNES) of Ti-L and C-K edges in rich layers of Ti and C, indicated in blue and red, respectively. The green profile is a reference to $Ti_3C_2T_x$ discussed in previous works.[10,42,44] ELNES is performed at the interface between the tribofilm and the substrate to verify if the MXene flakes have maintained their structure. The analysis reveals that the overall intensity of the $\pi^*$ peak in the C-K edge is less pronounced than the reference; however, the $\pi^*$ peak is still well defined, suggesting that the Van der Waals MXene structure is well preserved at this interface. Instead, the broad $\sigma^*$ peak may indicate loss of crystallinity, weakening of the Ti-C bonding, or oxidation for the MXenes at the interface with the substrate.[44,47,48] At the same relative scale as the C-K edges, the measured Ti edges are smaller in intensity than the reference. More importantly, the ELNES peaks appear left-shifted (see red arrow in Figure 5d), which is associated to a reduction of the oxidative state of $Ti^{3+}$ with respect to the reference.[49]

A semi-quantitative EELS analysis was performed by acquiring a map at the interface of interest, as illustrated in **Figure S11**. The colorized map is shown as an insert in Figure 5d, with the color code or red, green, and blue for Ti, C, and O, respectively. We extracted the elemental composition profile along the central line of the map (white dashed line in the color map). The atomic percentage profile following the same color code along the selected interface (~10 nm) is shown as an insert in Figure 5d. The semi-quantitative analysis revealed a relative composition of $Ti_3C_{1.5}O_{2.6}$, assuming a Ti oxidation state of 3+. The results show that Ti and C alternate at the interface, as expected, which is important for addressing the question of whether crystallinity is maintained. However, the analysis remains qualitative because surface terminations other than O are not considered and the oxidation state of Ti is constrained to 3+.

To summarize, TEM investigation including high-resolution imaging and interlayer spacing evaluation, SAED pattern calculation, and EELS analysis confirmed that MXene flakes coated by interfacial assembly deposition preserve their crystallinity after nanomechanical stress.

## 3. Materials and methods.

Sodium tetrafluoroborate ($NaBF_4$, ≥ 97 %), hydrochloric acid (HCl, 35 %) and dimethyl sulphoxide (DMSO, ≥ 99 %) were purchased from Carl Roth GmbH + Co. KG (Germany). Toluene (99+%) was purchased from Chem-Lab NV (Belgium). $Ti_3AlC_2$ MAX-phase powder (≤ 40 µm) was purchased from Carbon-Ukraine (Y-carbon LLC) (Ukraine).

### 3.1 ISA-MXene synthesis and thin film preparation

$Ti_3C_2T_x$ MXenes were chemically synthesized by selective etching of aluminium from $Ti_3AlC_2$ MAX-phase powder (≤ 40 µm particle size). Firstly, around 1g of MAX phase powder is stirred in 9 M HC for 18 hours to remove all metallic residues, vacuum-filtered on 0.45 µm cellulose-nitrate (CN) filter, rinsed with deionized (DI) water, and transferred in glassy carbon beaker.[10] For the chemical etching, 200 ml solution was prepared with concentrations of 1 M $NaBF_4$, 1 M HCl and 0.2 M DMSO, and added to the MAX phase and covered with parafilm.[50] Etching solution was added to max powder and stirred for seven days at 45 °C. After the etching was done, the powder was filtered on 0.45 µm CN filter and rinsed several times with DI water. In addition, powder was stirred and washed in 40 °C DI water and again filtered and dried in vacuum. Dried

MXene powder was transferred in 30 ml DMSO and stirred for 24 hours to intercalate DMSO molecules between layered flakes. When intercalation was done, DMSO was removed using a centrifuge at 13,000 rpm several times and replaced with DI water. Powder was transferred to 50 ml DI water and delaminated using ultrasonication for 1 h. Delaminated MXene powder was filtered and dried in vacuum before thin film preparation.

Thin-film preparation was done on freshly cleaved Mica using modified interfacial deposition.[32] Firstly, 0.5 mg/ml dispersion of MXenes in DI water was prepared by 1 h ultrasonication. The dispersion was centrifuged for 15 min at 3500 rpm and supernatant with fine MXene flakes was removed for further usage. 1 ml of supernatant is added to a mixture of 25 ml DI water and 2 ml toluene and stirred in a beaker for 20min. The mixture was added to a 20 ml DI water with mica placed on the bottom of the beaker. After few minutes, toluene layer got separated from the water, and MXene flakes have mainly been at the interface between two phases. The substrate was slowly lifted several times upwards with the tweezer, while keeping the surfaces parallel to the interface, to bind the MXene flakes. Six and ten dips were performed to prepare the AFM and SFA samples, respectively.

### 3.2 Atomic Force Microscopy

**3.2.1 General information and calibration.** AFM measurements were performed with an AFM Cypher Asylum (Oxford Instruments). A freshly cleaved mica sheet was glued on a metal support to act as substrate on which to transfer the MXene flake by interlayer deposition. A CONT GB – G gold-coated tip (BudgetSensors) was employed for the AFM investigation. Characteristics of the tip: Nominal tip radius $R < 25$ nm, measured resonance frequency $f_0 = 12.793$ KHz from thermal fit, nominal length $L = 450\ \mu m$, nominal width $w = 50\ \mu m$. For data analysis and interpretation, an effective tip radius $R^* = 26\,nm$ is considered to compensate for plastic deformation effects of the gold probe with respect to the nominal value $R$. The tip elastic constant in the normal direction is obtained from the Sader method (Q=74.807) as $k_{norm} = 0.24\,\mathrm{N/m}$ given the density of air at room temperature. The lateral elastic constant is obtained from a torsion-based lateral stiffness calculation given the nominal height ($h = 17\,\mu m$) from the tip to the cantilever neutral plane: $k_{lat} = k_{norm} * \frac{4G_{Si}}{3E_{Si}}\left(\frac{L}{h}\right)^2 = 89\,\frac{\mathrm{N}}{\mathrm{m}}$ where $G_{Si} = 60\,\mathrm{GPa}$ and $E_{Si} = 150\,\mathrm{GPa}$ are the shear and elastic moduli of Silicon respectively. The bulk parameters of Silicon are selected over gold since the torsion analysis is performed away from contact with the substrate. In this case, the material composition of Silicon dominates (93%), thus, we assumed that the Au coating was not affecting the torsion performance of the tip, namely $k_{lat}$.

Forces in nN are extracted from voltage read-outs by multiplying $k_{norm}$ by the measured DeflInvols (249.96 nm/V, calculated via thermal calibration) for the normal composition of the force, and by multiplying DeflInvols by $k_{lat}$ and the ratio h/L for the lateral composition of the force.

**3.2.2 Adhesion and Friction maps.** The adhesion map was obtained in contact mode AFM with a force scan rate of 0.99 Hz, and a scan speed (lateral tip velocity) of 7.87 µm/s.

The friction map was obtained using the software Gwyddion to perform arithmetic on the lateral signal: $F_f = (Trace - Retrace)/2$. The arithmetic process was performed to remove the topography-couple component recorded on the Lateral Deflection signal during acquisition. The result of this operation is a 2D friction force map where each line represents a 1D friction profile in the (x,y) spatial coordinates.

**3.2.3 Effective elastic modulus for the area calculation in DMT model.** A calculation on the effective elastic modulus at the mica/MXene/Au interface was done to derive a sensible value of the contact area $A(x,y)$ in the DMT model. Table 1 lists the different materials and interfaces in terms of their elastic modulus and Poisson's ratio with references.

*Table 1 Mechanical characteristics of the mica/MXene/gold interface from literature.*

| Material | Elastic modulus $E$ (GPa) | Poisson's ratio $\nu$ |
|---|---|---|
| Au | 78[36] | 0.44[36] |
| Mica | 170[51] | 0.25[51] |
| $Ti_3C_2T_x$ | (in plane): 330[16] – 484[17] | 0.20 (DFT calculation)[52] |
| $Ti_3C_2T_x$ | (out-of-plane) 80-100[53] | |

The in-plane elastic modulus has been typically selected for nanomechanical studies of $Ti_3C_2T_x$, however, recently the out-of-plane composition of the elastic modulus component perpendicular to the basal plane was derived by combining in situ electron microscopy with AFM nanomechanical mapping. Thus, we selected $E_{MXene} = 90$ GPa and $\nu_{MXene} = 0.20$ for our calculation. Then, we calculated the effective elastic moduli for each interface with respect to the gold probe: $E^*_{i,Au} = \left(\frac{1-\nu_i^2}{E_i} + \frac{1-\nu_{Au}^2}{E_{Au}}\right)^{-1}$ which gives the following values: $E^*_{\mathrm{Ti3C2Tx/Au}}$= 47.6 GPa; $E^*_{\mathrm{mica/Au}} = 63.1$ GPa. For simplicity, in the discussion, the effective elastic modulus is labeled without the reference to the probing material (Au): $E^*_{\mathrm{Ti3C2Tx}}$= 47.6 GPa.

### 3.3Transmission Electron Microscopy

TEM analysis was performed at USTEM (TU Wien). TEM lamellae were prepared via focused ion beam (FIB) milling using a ThermoFisher Scios II system. Microstructural characterization was carried out on an FEI Tecnai F20 transmission electron microscope equipped with an X-FEG field emission gun and operated at 200 kV. TEM micrographs were recorded with a Gatan Rio16 camera. Chemical mapping and composition profiles were acquired via energy-dispersive X-ray spectroscopy (EDS) using an EDAX-AMETEK Apollo XLTW SSD detector, while electron energy-loss spectroscopy (EELS) was conducted using a Gatan GIF Tridiem spectrometer.

### 3.4 Surface Forces Apparatus

A home-built surface forces apparatus (SFA) set-up was used as in our previous works. [54–56] For the back-silvered mica surfaces, 35 nm Silver was deposited using physical vapor deposition (PVD) at a pressure of 1.7 × 10−6 mbar (system built at TU Wien). The mica layer is secured onto a cylindrical quartz disk (from SurForce LLC.), with a radius of curvature $R_C$ = 0.02 m, by applying UV-cured NOA 81 glue. SFA can measure thin films thickness based on multiple beam interferometry. Briefly, the light from the optical cavity defined by the two facing silver mirrors (i.e., back-silvered mica surfaces), produces an interference pattern (Newton's rings) and fringes of equal chromatic order (FECO) via a diffraction grating. FECO are projected onto a 2D detector in the spectrometer. By placing the two back-silvered mica surfaces into contact, the thickness of the mica layers is defined as the spacing between the mirrors ($T_0$). When the MXenes are coated onto the mica (Section 3.1) and the surfaces are brought again into contact, the MXene thin film will change the spectrum, which we fitted by introducing a new thickness $T$, as illustrated in Figure S2. The analysis of the FECO spectra is performed with SFAExplorer.[54–56]

## 4. Conclusions

In this work, we studied the nanomechanical compliance of interfacial self-assembled MXene flakes (i.e., $Ti_3C_2T_x$) against a gold-coated AFM tip. For the first time, we quantified the interfacial shear stress at the $Ti_3C_2T_x$ /Au interface by performing lateral spectroscopy AFM. Furthermore, we confirmed that the flakes' crystallinity is maintained within the tribofilm via a systematic TEM investigation. The main results are summarized below:

- HR-AFM topography and SFA confirmed the expected MXene thickness after the interfacial assembly deposition to be ~ 1.2 nm, and a strong resilience and stability in ambient conditions over time (min. four months).
- A combination of FM-AFM and LS-AFM provided insights into the nanomechanical response of MXene flakes against a gold probe, which resulted in the derivation of an average interfacial shear stress for MXene of ~399 MPa. In addition, HR-AFM showed the presence of nano-wear following the LS-AFM tests, resulting in a reduction of the flake thickness to ~0.65 nm (central region of the flake).
- HR-TEM imaging, EDX, and SAED confirmed the chemical and crystalline structure of MXene in different areas of the tribofilms. Moreover, EELS showed that the flake retained the typical characteristics of $Ti_3C_2T_x$ at the interface to the substrate.

In conclusion, the interfacial assembly method on mica has provided single MXene flakes which are ideal platforms to perform nanomechanical characterization. The nanomechanical quantification presented here against a metal counterbody is crucial for application in nanotechnology, paving the road for a better understanding of MXenes at metal contact, with large scope for novel tribovoltaic nanogenerator devices, electric vehicles, and robotics.

## 5. Associated content

SUPPORTING INFORMATION

The supporting information is available free of charge at the ACS Publications website.

AUTHOR INFORMATION

**Corresponding Author**

*Pierluigi Bilotto

pierluigi.bilotto@tuwien.ac.at

**AUTHOR CONTRIBUTIONS**

PB: Conceptualization, Methodology, Validation, Formal analysis (AFM, SFA, HRTEM, EELS), Investigation (AFM, SFA), Data curation, Writing (Original draft, Review, and Editing). SS: Validation, Formal Analysis, Investigation (HRTEM, SAED), Writing (Original draft, Review, and Editing). MP: Investigation (MXene synthesis, Interfacial assembly coating), Writing (Original draft, Review, and Editing). IP: Investigation (SFA), Writing (Original draft, Review, and Editing). MSP: Formal Analysis (EELS), Investigation (EELS), Writing (Review, and Editing). CG: Writing (Original draft, Review, and Editing) The manuscript was written through contributions of all authors. All authors have given approval to the final version of the manuscript.

ACKNOWLEDGMENT

Pierluigi Bilotto would like to acknowledge Prof. Markus Valtiner (TU Wien) for granting accessibility to his laboratories at the Institute of Applied Physics (E134-02). The authors thank the TU Wien Library for its financial support through the Open Access Publishing Fund. Part of this work was carried out as part of the COMET Centre InTribology (FFG no. 906860), a project of the “Excellence Centre for Tribology” (AC2T research GmbH). In Tribology is funded within the COMET – Competence Centres for Excellent Technologies Programme by the federal ministries BMIMI and BMWET as well as the federal states of Niederösterreich and Vorarlberg based on financial support from the project partners involved. COMET is managed by The Austrian Research Promotion Agency (FFG). The authors acknowledge ”Gesellschaft für Forschungsförderung Niederösterreich m.b.H.” through its FTI PhD Funding Programme (FTI22-D-018).

# TOC

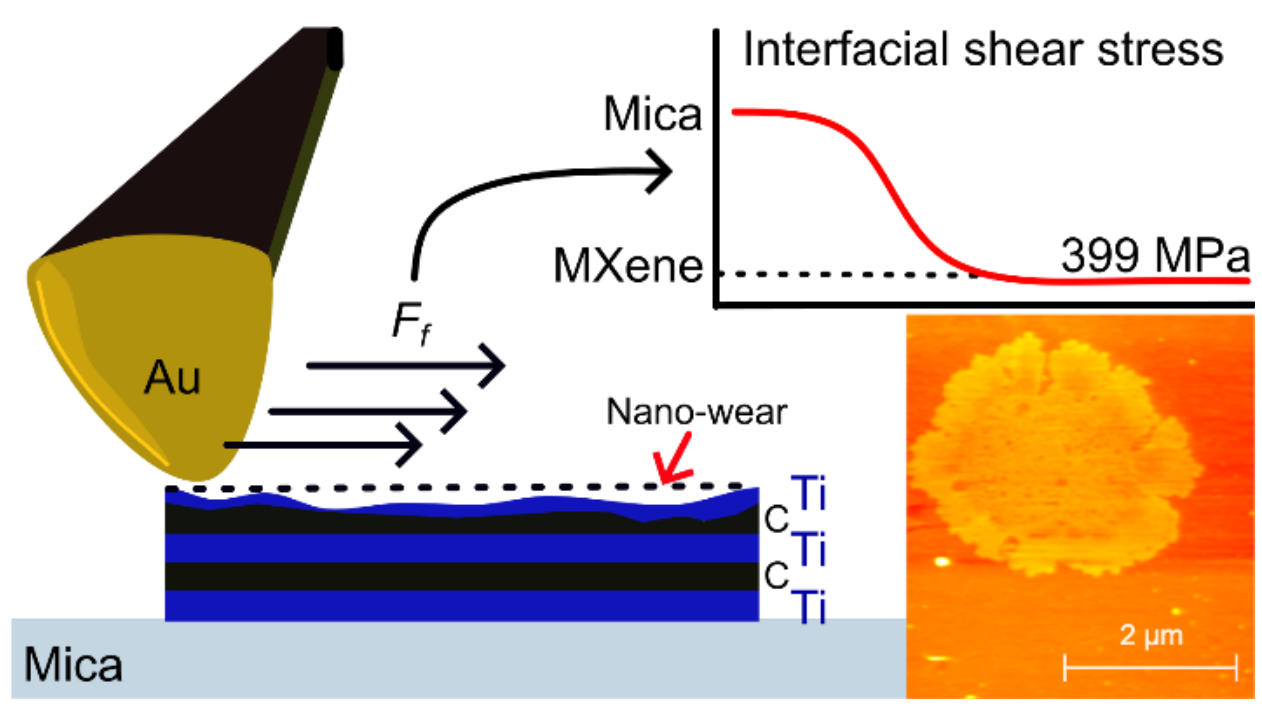

Supportive information

# Nanomechanics of MXene flakes at gold interfaces

*Pierluigi Bilotto[1]*, Sabine Schwarz[2], Marko Piljevic[3], Iago Peters[4], Michael Stöger-Pollach[2], Carsten Gachot[1]*

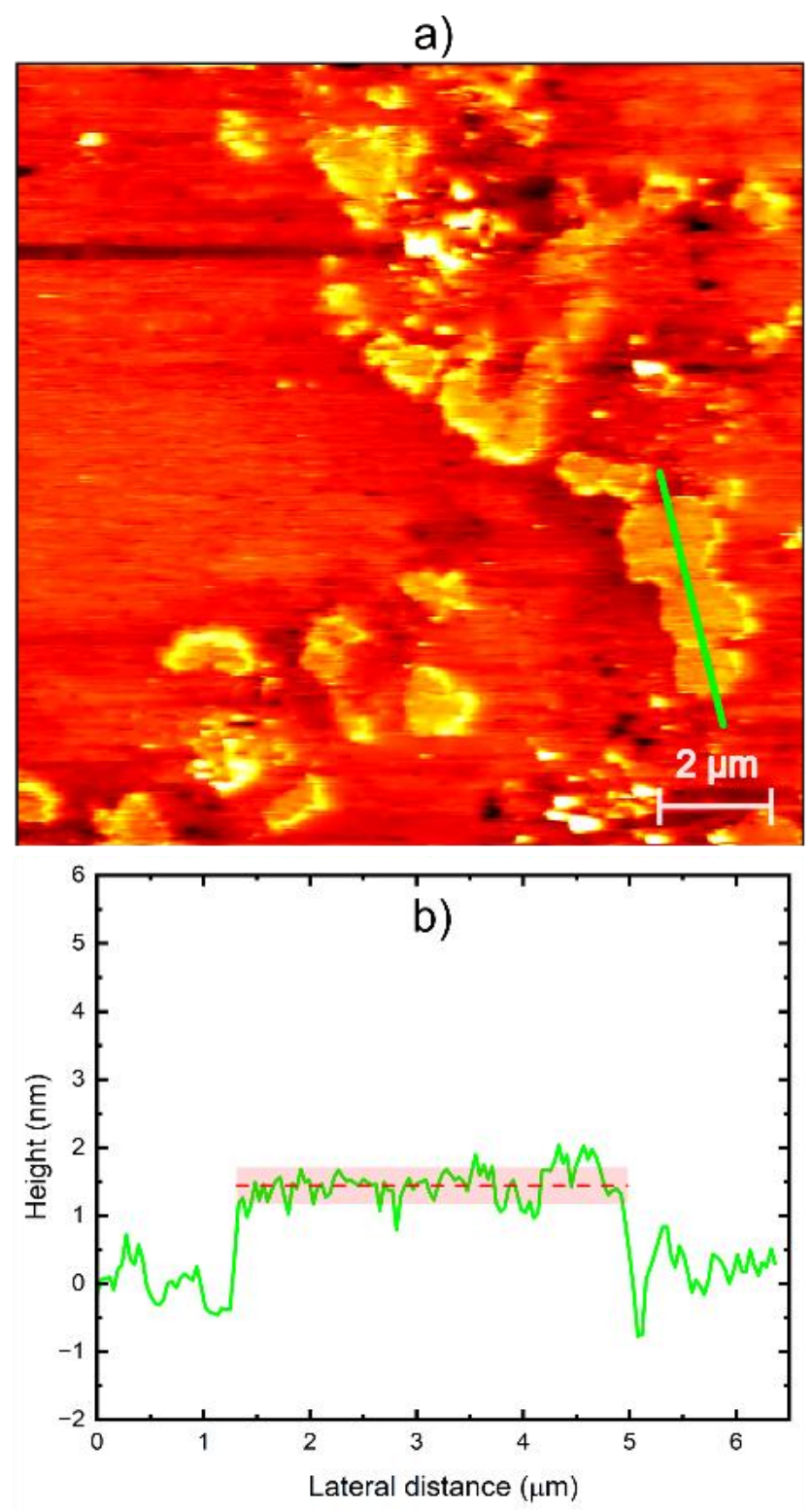

*Figure S 1: As deposited ISA-MXene thin film on large scan size (14 x 14 um).*

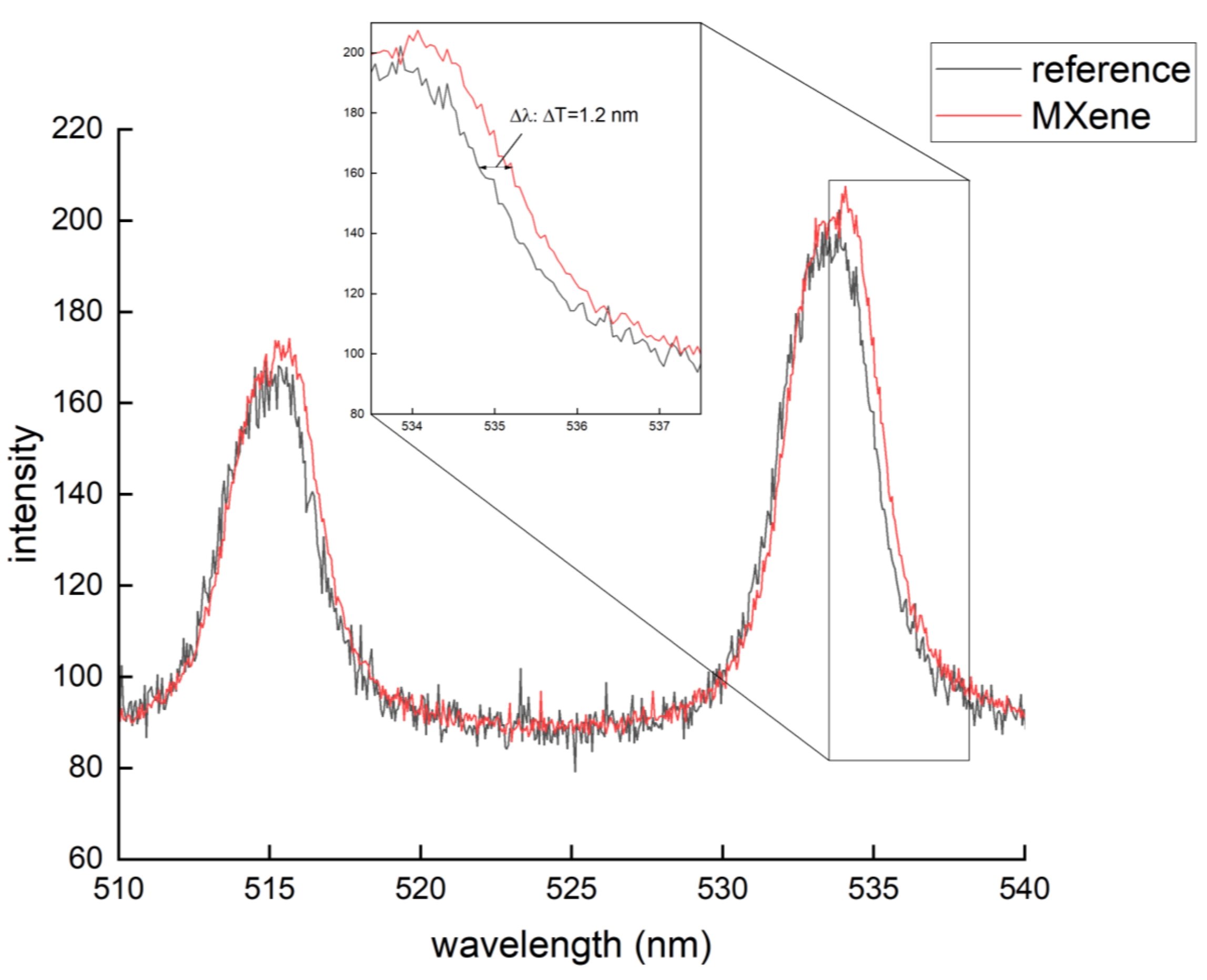

*Figure S 2: Wavelength spectrum from FECOs in SFA. In black is the reference, i.e., mica, and in red the MXene. The MXene flakes generate a shift in the spectrum which could be fitted with SFAExplorer to give the MXene thickness.*

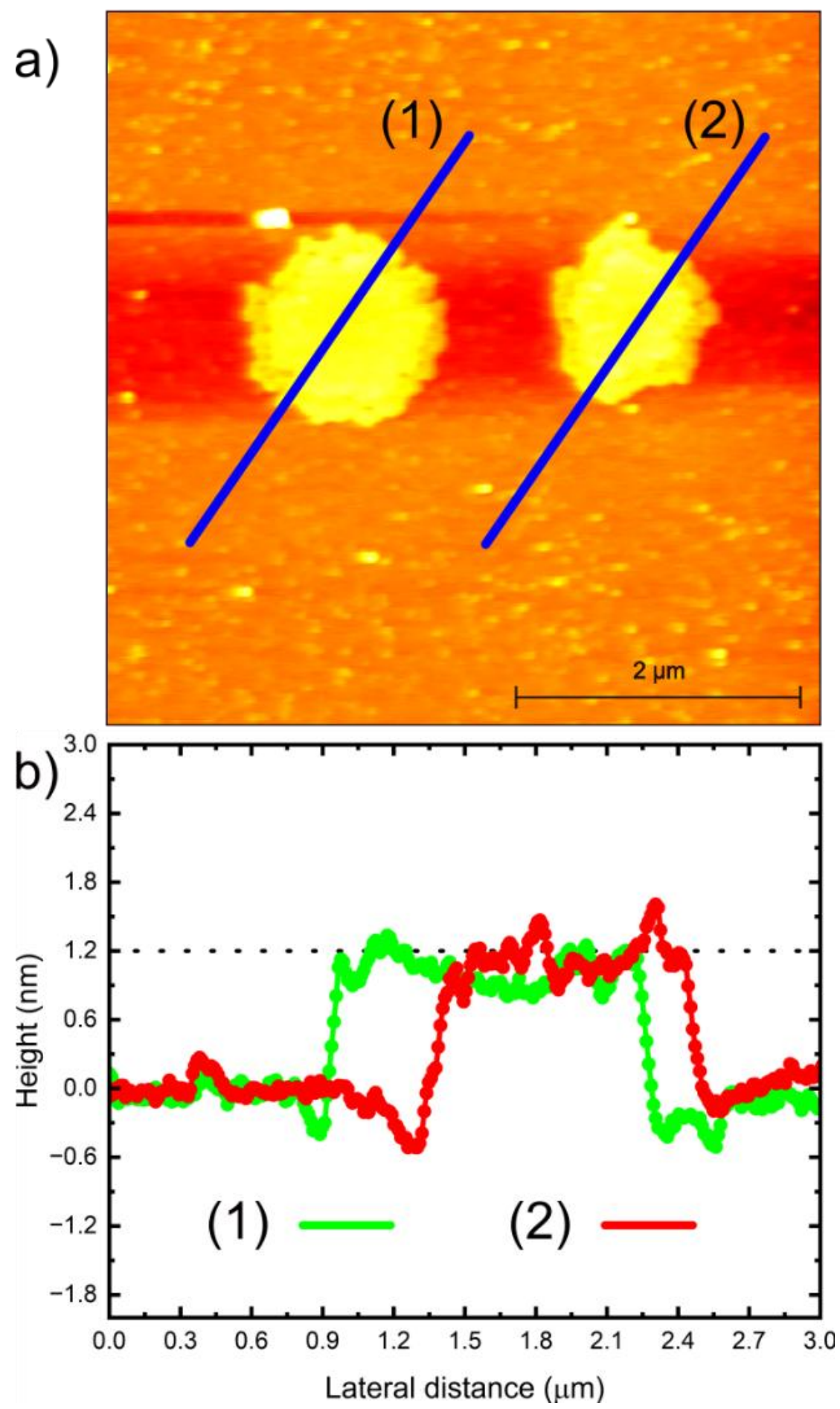

*Figure S 3: AFM topography from the sample shown in Figure 1. The sample was stored for four months at room temperature. The height profiles are consistent with the one shown in Figure 1.*

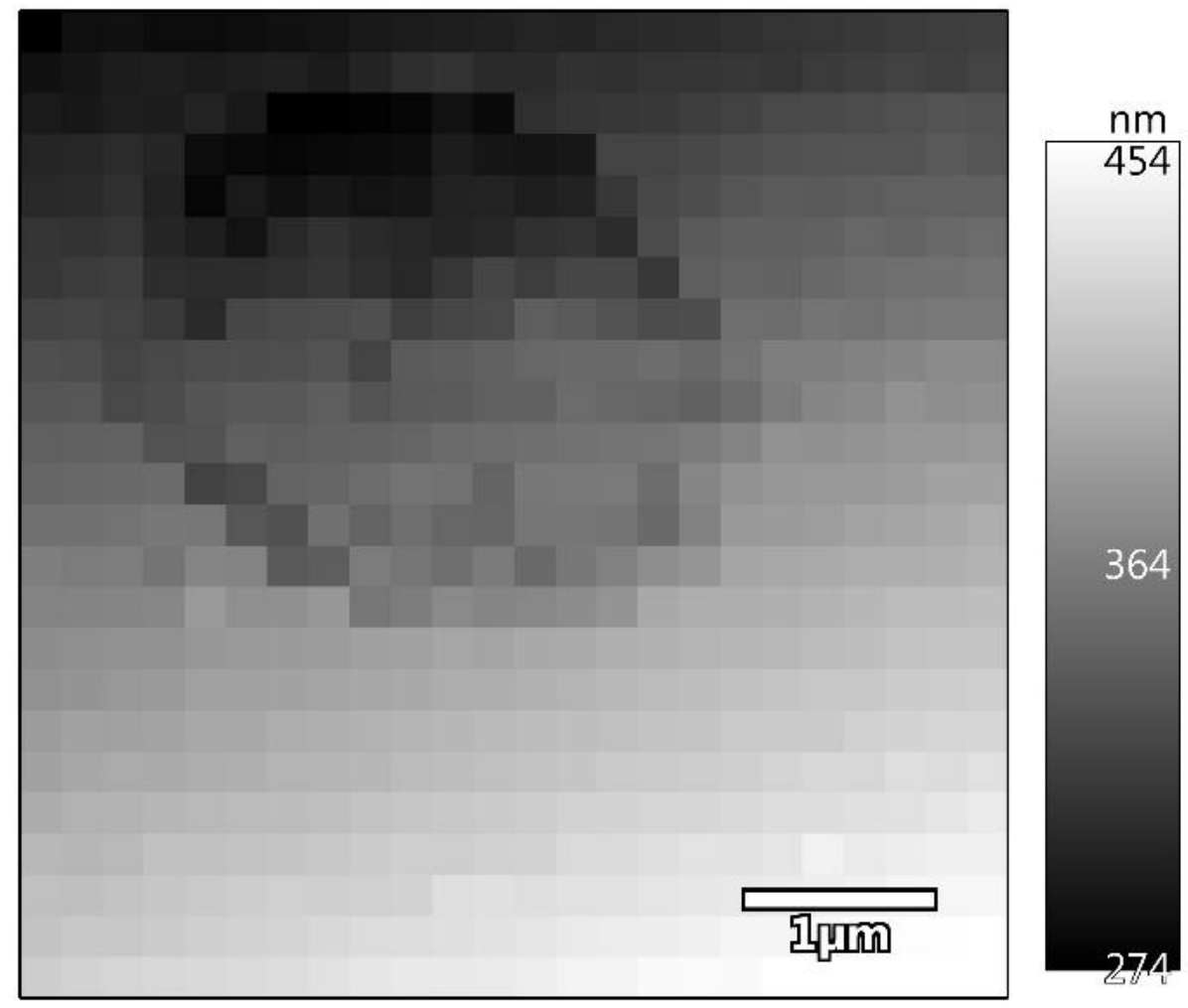

*Figure S 4: Topography image obtained from the force map acquired on F.7.*

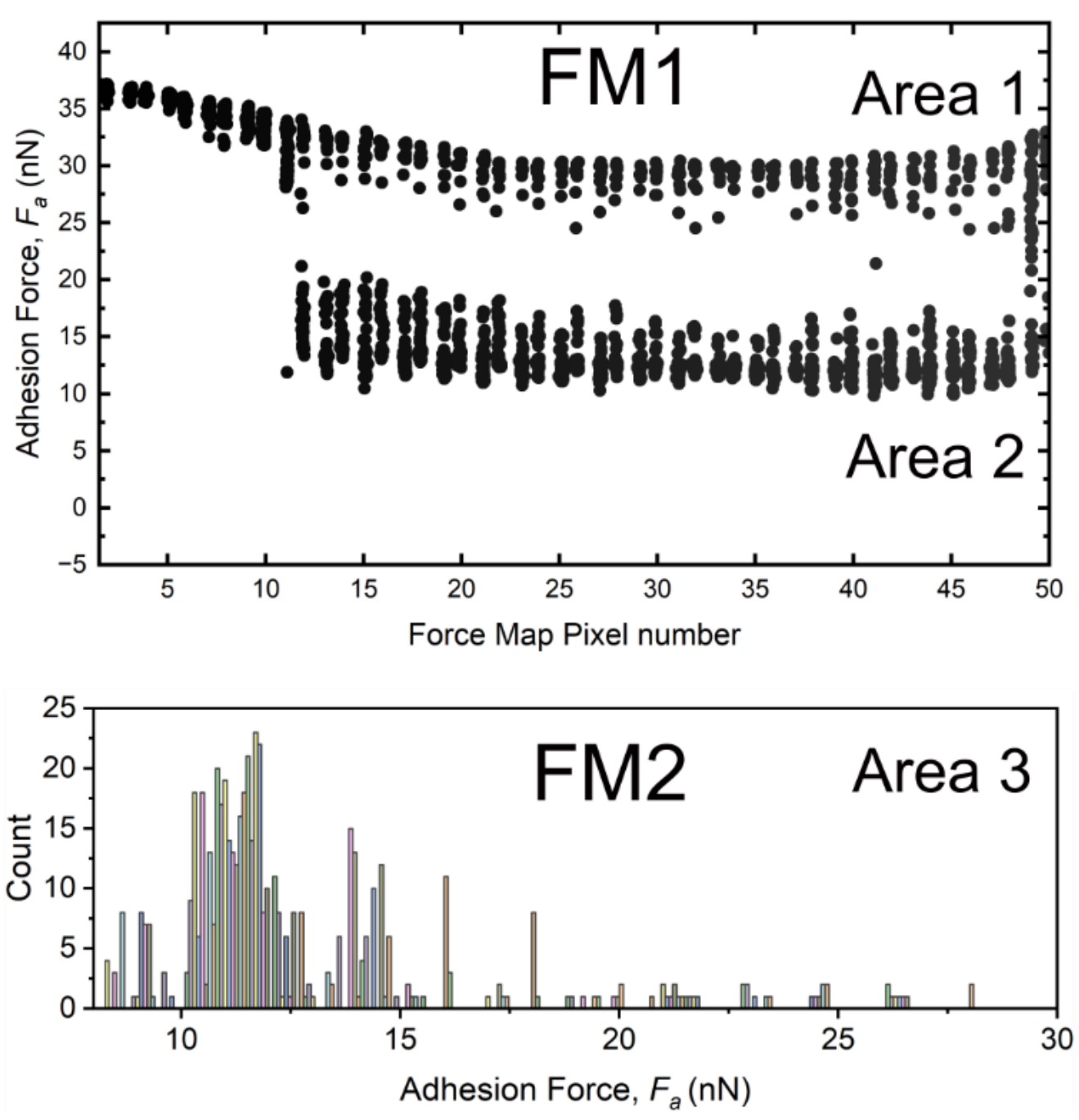

*Figure S 5: Comparison of adhesion forces measured out of two force maps (FM). Note that the values are distributed between two areas representing MXene flake (Area 1 and Area 3) and mica (Area 2).*

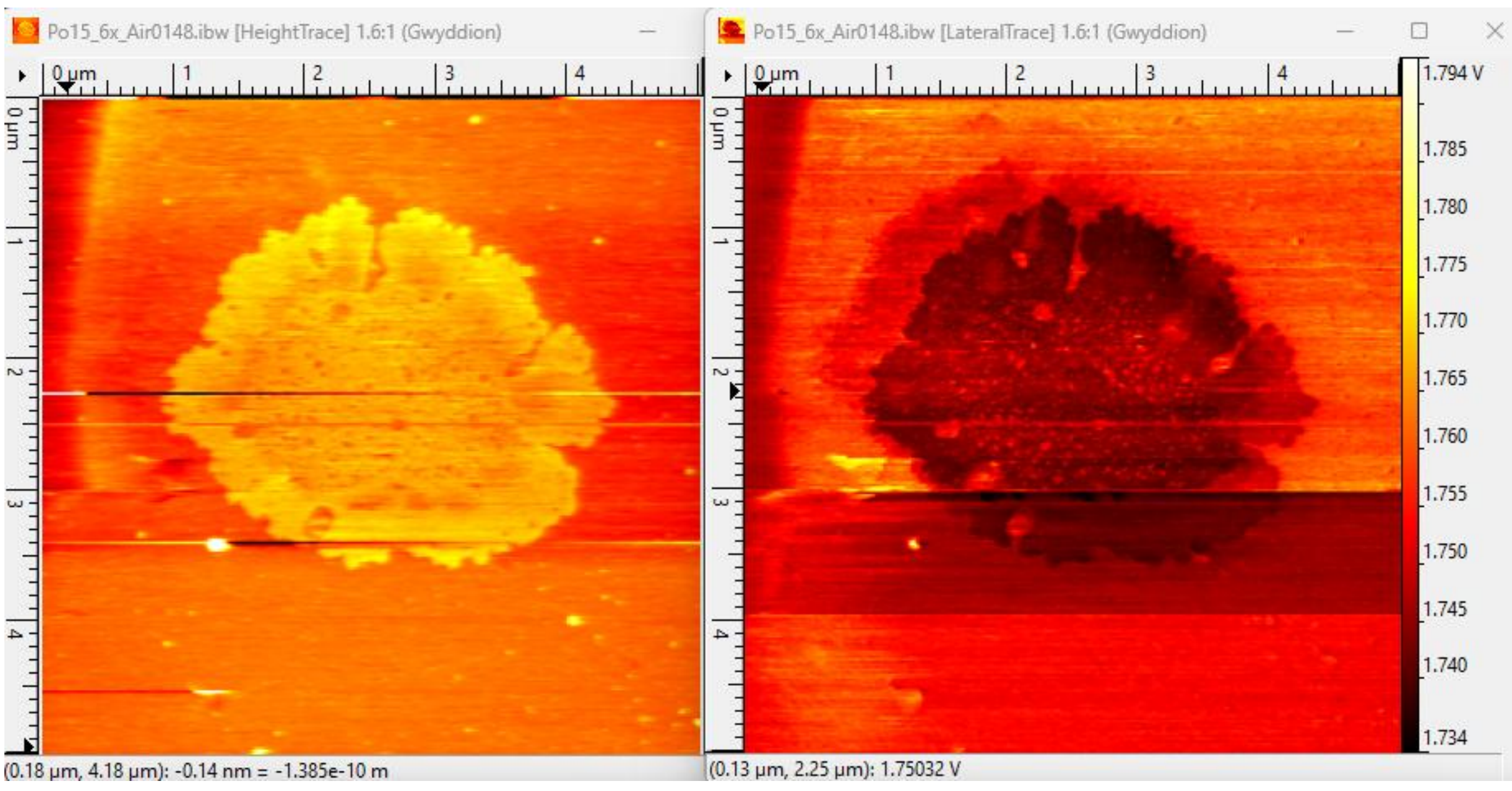

*Figure S 6: Lateral trace and retrace of the flake*

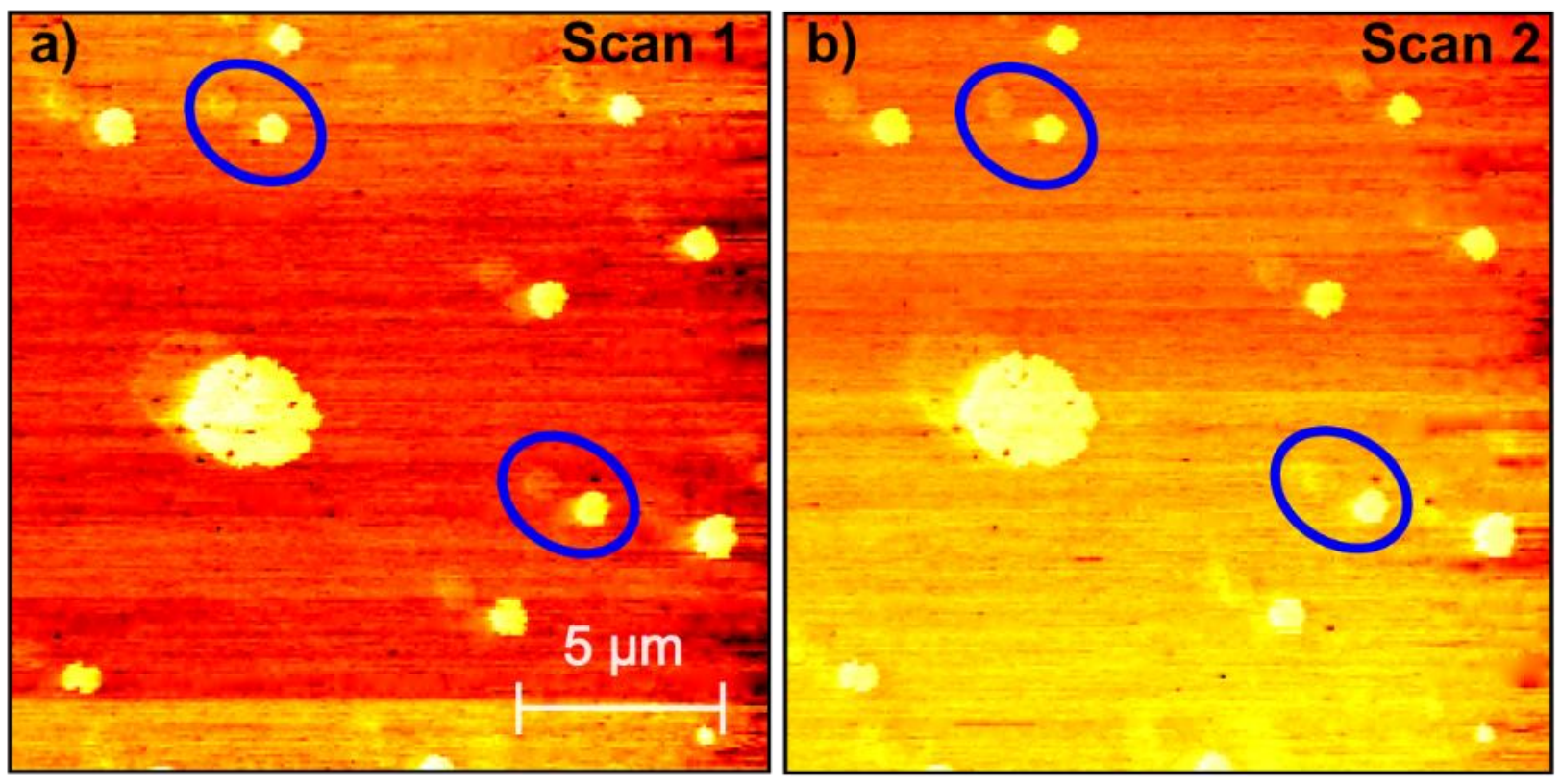

*Figure S 7: Analysis of shadows observed during AFM topography*

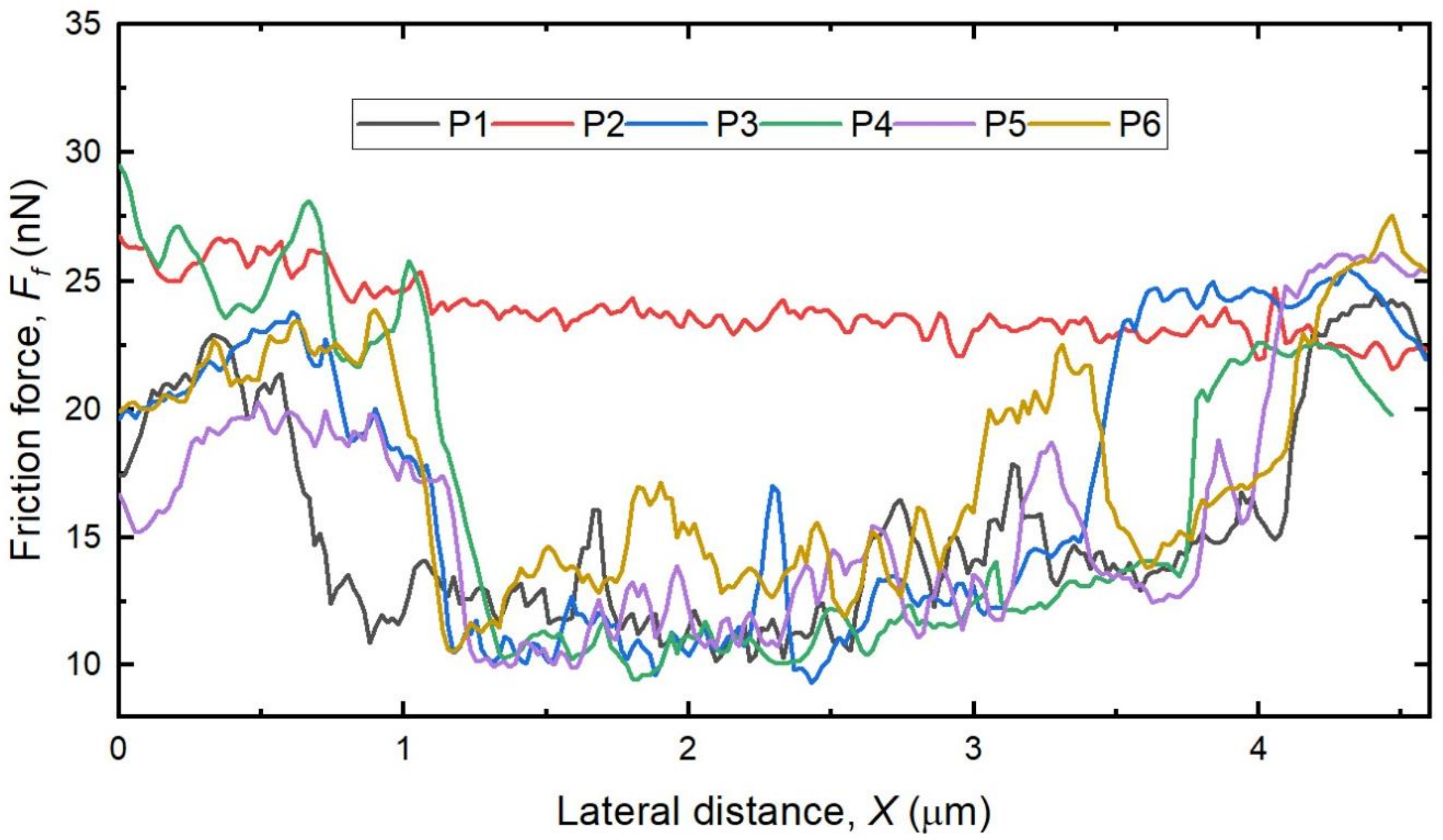

*Figure S 8: Friction force plotted as a function of the lateral distance. The section points are the same defined in Figure 3..*

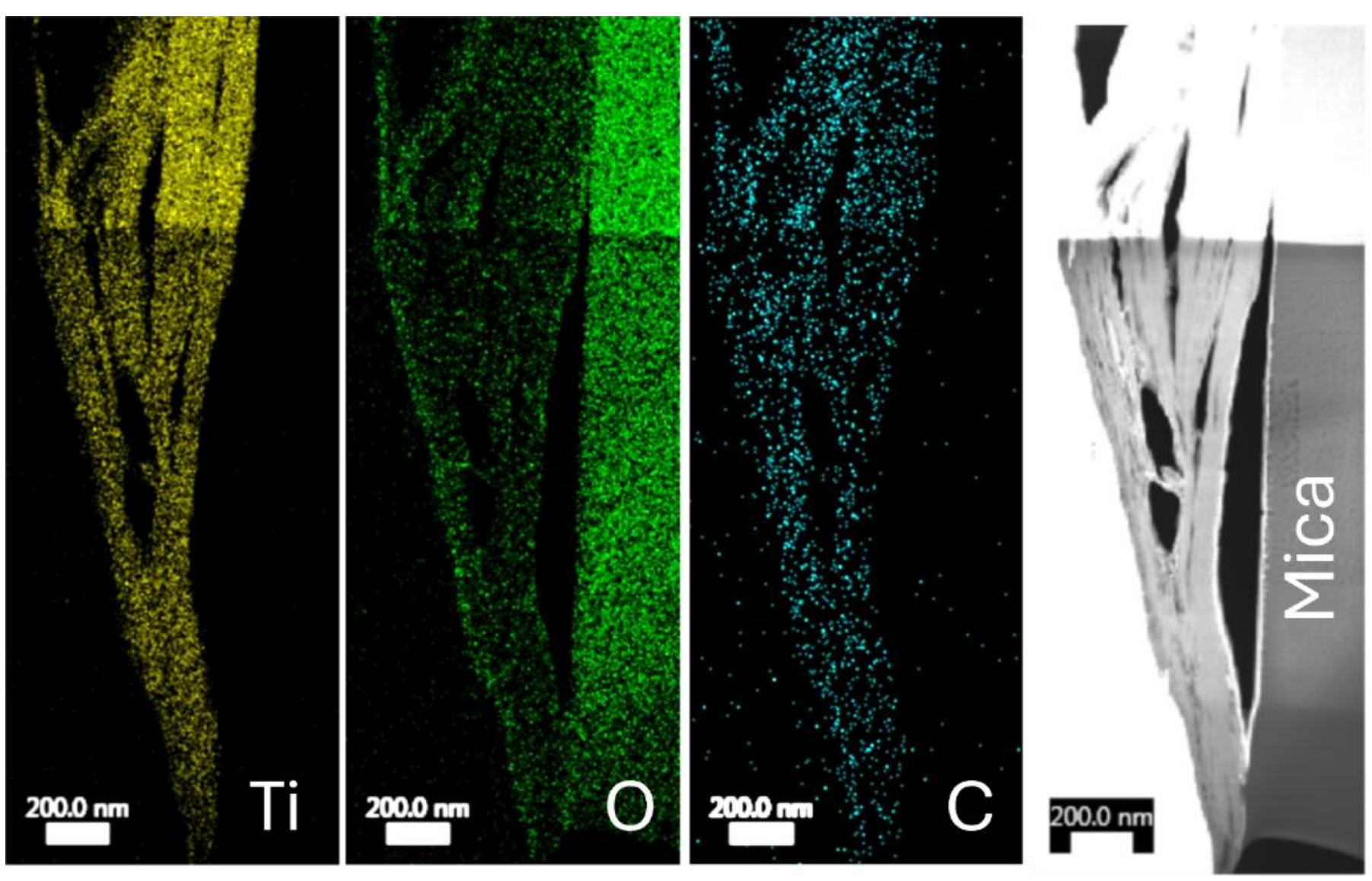

*Figure S 9: EDX mapping of the FIB lamella*

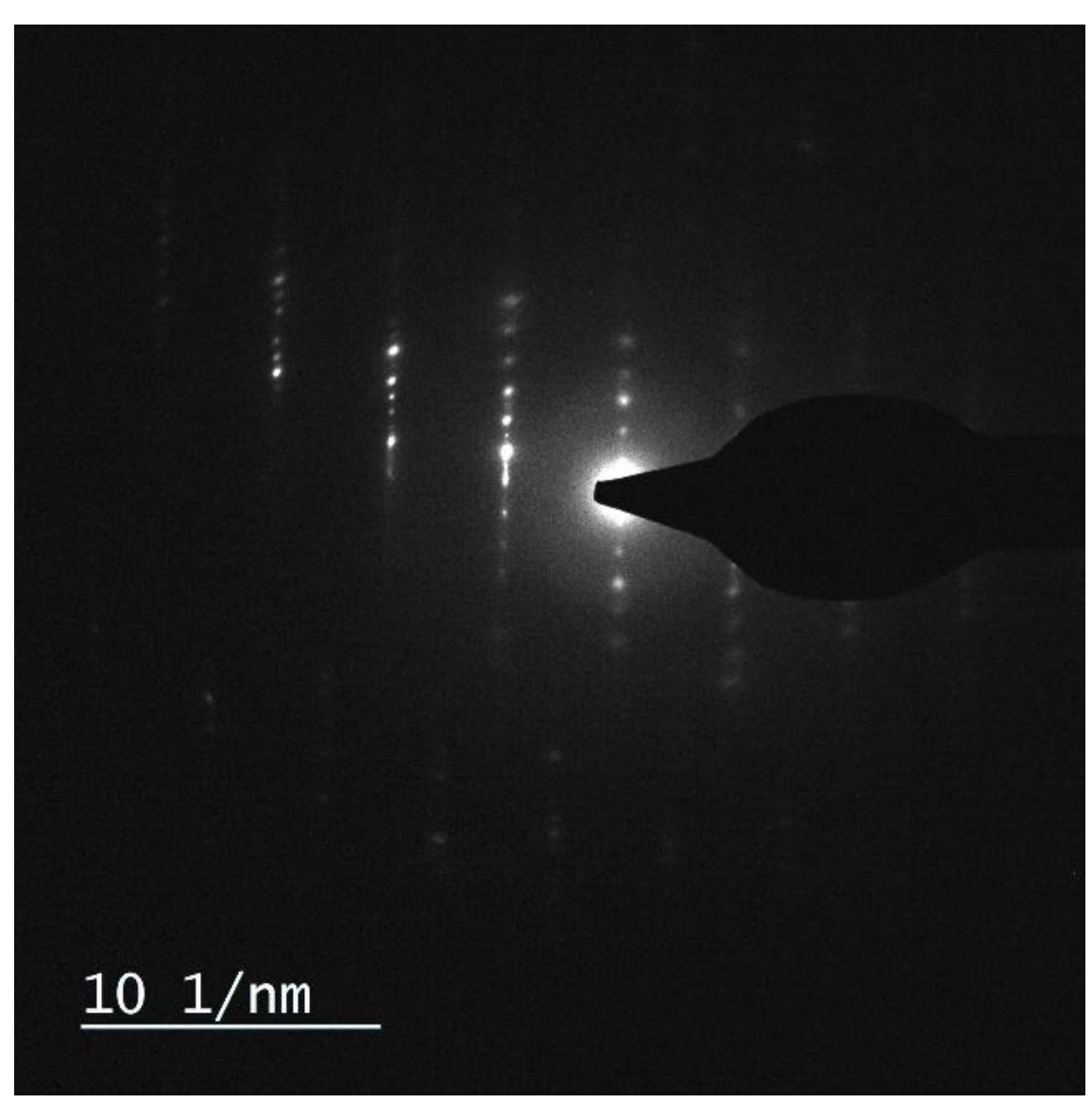

*Figure S 10: SAED of mica*

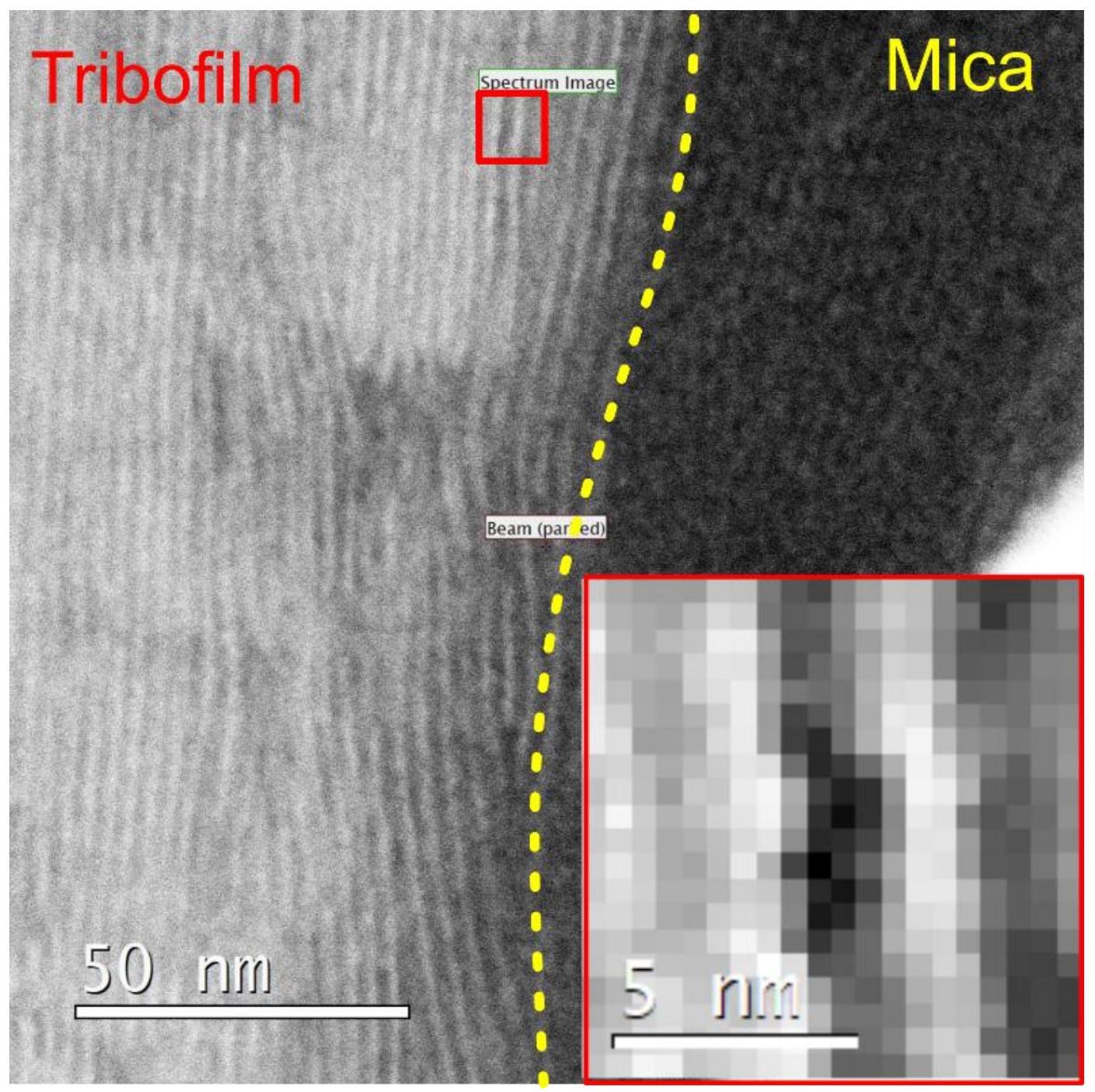

*Figure S 11: Detail of the area where EELS is performed at the interface with the substrate (yellow dashed line). The red square is the selected area in which is perfomed the EELS map colored in Figure 5d.*